\documentclass[aps,prd,groupedaddress,preprintnumbers,twocolumn]{revtex4-1}

\usepackage{graphicx}%
\usepackage{amsmath,amssymb,amsfonts}%
\usepackage{xcolor}%
\usepackage{tabularx,ragged2e,booktabs,subfigure}
\usepackage[shortlabels]{enumitem}

\newcommand{\DURaff}{Institute for Particle Physics Phenomenology, University
  of Durham, South Road, Durham DH1 3LE, UK}
\newcommand{\CERNaff}{Theoretical Physics Department, CERN, 1211 Geneva 23, Switzerland}

\newcommand{\IFAEaff}{Institut de Física d’Altes Energies (IFAE), The Barcelona Institute of Science and Technology, Campus UAB, 08193 Bellaterra (Barcelona), Spain}
\newcommand{\UABaff}{Grup de Física Teòrica, Dept. Física,
Universitat Autònoma de Barcelona, E-08193 Bellaterra, Barcelona, Spain}

\newcommand{\MADaff}{Departamento de F\'isica Teorica C-15 and CIAFF, Universidad Aut\'onoma de Madrid, Madrid, Spain}

\begin{document}

\preprint{CERN-TH-2024-200, IPPP/24/73}

\title{A Cell Resampler study of Negative Weights in Multi-jet Merged
  Samples}


\author{Jeppe R.~Andersen}       \affiliation{\DURaff,\CERNaff}%
\author{Ana Cueto} \affiliation{\MADaff}
\author{Stephen P. Jones}       \affiliation{\DURaff}%
\author{Andreas Maier}       \affiliation{\IFAEaff}\affiliation{\UABaff}%


\date{\today}

\begin{abstract}
We study the use of cell resampling to reduce the fraction of negatively weighted Monte Carlo events in a generated sample typical of that used in experimental analyses.
To this end, we apply the Cell Resampler to a set of $pp \rightarrow \gamma \gamma + \mathrm{jets}$ shower-merged NLO matched events, describing the diphoton background to Higgs boson production, generated using the FxFx and MEPS@NLO merging procedures and showered using the Pythia and Sherpa parton shower algorithms.
We discuss the impact on various kinematic distributions.
\end{abstract}


\maketitle

\section{Introduction}

Theoretical predictions are required at ever higher precision in order to match the accuracy of the experimental analyses from LHC.
The generation, storage and processing of these theoretical simulations are increasingly becoming a computational bottleneck for the experimental collaborations~\cite{CERN-LHCC-2020-015}.
The hard, perturbative fixed order calculations at N$^m$LO, with $m\ge1$, are formulated as explicit integrals over phase spaces of varying multiplicity.
The contribution from the sampled phase space configurations are stored as \emph{events} with the kinematic information and a \emph{weight}, which reflects the evaluation of the integrand for the particular phase space point.
Higher order calculations typically contain configurations with both positive and negative weights.
The measured cross sections are by definition non-negative, and although the theoretical predictions contain contributions with both signs, the expectation is that for physically reasonable predictions the sum will be non-negative for all arbitrarily differential configurations.

The negative weights originate from both genuine interference terms as well as subtraction terms introduced to handle the cancellations of divergences between the integrals of varying multiplicity.
The cancellation of divergent contributions between phase spaces of different multiplicity is organised with subtraction terms and leaves large numerical cancellations between events in the divergent regions of phase space.
In addition, the Born-virtual interference events, which can be negative, are often generated separately from the Born contribution for practical reasons, leaving both positive and negatively weighted events in the Born phase space.
If an infra-red safe definition of the locality of an event is given, then it may be possible to organise the numerical cancellation of weights of opposite signs between nearby events.
This is the starting point for the Cell Resampler (CRES)~\cite{Andersen:2021mvw,Andersen:2023cku}.
The weight of events would then be correlated with the local contribution to the cross section, just as is the case at Born level.
Reducing the numerical magnitude of the cancellation between events will reduce the variance of the weights which has multiple beneficial effects.
Firstly, the reduced variance leads to a more efficient unweighting procedure.
Secondly, correlating the weight of the event with the contribution to the cross section means that the event weight is the optimal choice in the unweighting procedure, with the number of events in each phase space region reflecting the local contribution to the cross section rather than the degree of cancellation.
In other words, regions with small cross section will receive fewer events than regions with a larger cross sections.

Reducing the number of negative events is particularly useful in situations where the fixed order calculation is but the start of the computational complexity.
Examples of such include either the use of NLO events in parton shower matching~\cite{Frixione:2002ik,Frixione:2007vw} and merging~\cite{Frederix:2012ps,Gehrmann:2012yg,Lonnblad:2012ix} as well as detector simulation.
It also facilitates the reuse of computationally expensive fixed order calculation, for example the calculation of new kinematic distributions using stored N$^2$LO events~\cite{Czakon:2023hls}, which exhibit a far more intricate cancellation structure than NLO events.

The first issue in defining a local neighbourhood of an event is then to define a metric on the set of events.
We seek a metric which results in infra-red (IR) safe distances between events, and assigns a small distance between events when the momenta of their IR safe objects differ only slightly.
The quantities which should enter the metric should therefore be based on the momenta of clustered jets rather than the momenta of individual QCD partons.
Various metrics on the space of events have been studied elsewhere~\cite{Komiske:2019fks}.
We will here present results obtained with the absolute metric discussed in Ref.~\cite{Andersen:2021mvw,Andersen:2023cku}.

The Cell Resampler~\cite{Andersen:2021mvw,Andersen:2023cku} neutralises the effect of the negative event weights by identifying the closest neighbours to the events with the most negative weights.
For each of these events (a \emph{seed}) a set of events (a \emph{cell}) is defined by adding events with increasing distance to the cell until either the sum of weights in the cell is non-negative, or the pre-specified maximum allowed distance is reached.
The weight of each of the events in the cell is then set to the sum of weights divided by the number of events in the cell, such that the total sum (and therefore the cross section of the cell) is unchanged.
The maximum allowed distance ensures that even in sparsely populated regions of space (where the cross section may not have reached statistical significance and still be negative locally) or on samples with a very large distance between negative weighted events and sufficient events with a positive weight, the effect on kinematic distributions from averaging weights to achieve a positive contribution is still limited to be local.
A similar approach based directly on observables~\cite{Andersen:2020sjs} or neural network classifiers~\cite{Nachman:2020fff,Nachman:2021opi} instead of cells were studied elsewhere.

It was previously demonstrated that such cell resampling can significantly reduce the fraction of negative weighted events in NLO fixed order samples for processes such as $W+$jets and $Z+$jets~\cite{Andersen:2021mvw,Andersen:2023cku}.
In this work, we investigate the application of cell resampling techniques to a set of events from a significantly different type of calculation, namely NLO matched and multi-jet shower merged events. Such samples are used by experiments and are relevant when effects from both fixed order corrections and the parton shower are important.

These event samples differ in several ways compared to the fixed order samples previously studied using cell resampling.
Firstly, they consist of a merged set of showered event samples of increasing multiplicity, each matched to the relevant fixed order NLO accuracy.
Each of these multiplicities can itself contain events with negative weight.
The multi-jet merging procedure further modifies the weights, here we consider both the FxFx~\cite{Frederix:2012ps} and MEPS@NLO~\cite{Gehrmann:2012yg} merging procedures.
Secondly, the kinematics of the events is significantly more complicated due to the parton showering generating further partons.
Finally, the events are sometimes unweighted, this means that the absolute value of the weight of each event in the sample is constant, however, the sample can still contain a large number of negative weight events.

A simulation that produces unweighted events with both positive and negative signs can attain the same statistical accuracy as one that produces events with only a positive weight by generating a number of events which is larger by a factor
\begin{align}
  \label{eq:cr}
 c(r_-) = \frac 1 {(1-2 r_-)^2 },
\end{align}
where $r_-$ is the fraction of negatively weighted events~\cite{Frederix:2020trv} (neglecting correlations between the negative and positive events).

The primary goal of our study is two-fold. First to establish whether the method of the Cell Resampler will work at all in a setup with events generated using a multi-jet merged method.
Such methods apply a mix of definitions of IR safe quantities such as jet definitions and photon isolation in the merging procedure, which can vary significantly from that used in the analysis.
Previous applications of the cell resampler have focused on NLO calculations, where the jet definitions of the calculations are tailored to a specific analysis.
The cell resampling method is particularly useful if the results are agnostic of a given analysis.
We will demonstrate that this quality is unmodified, as long as the input to the metric of the cell resampler uses the most resolving IR-safe definitions used in the generation and analysis.

In this current study we implement support for reconstructing the momenta of isolated photons within CRES, and apply it to event samples of the diphoton background to Higgs boson production.
The infra-red safe objects entering the metric are therefore not just the jets found in jet-clustering, but also photons.
We investigate the impact on the predictions for a published ATLAS analysis~\cite{ATLAS:2021mbt}.
In particular, we will discuss the impact on a) observables from momenta which directly enter the metric, b) observables from derived momenta and c) the distribution of the cross section on weights both positive and negative.
We will illustrate the impact using event samples generated with two different NLO merging algorithms.

In Section~\ref{sec:metr-cell-resampl} we describe the metric used with the Cell Resampler in the current study.
The metric is both a distance measure and a description of which perturbative objects should be included in this measure.
Section~\ref{sec:results} is devoted to an investigation of the Cell Resampler on the NLO accurate shower-merged event samples.
The amount of cancellation allowed will always be a trade-off between allowing large cells, which could change kinematic distributions but implement large cancellations, and requiring small cells with fewer events to implement cancellations between but also an acceptable (if any) change in the kinematic distributions.
Our conclusions are presented in Section~\ref{sec:conclusions}.

\section{The Metric of the Cell Resampler}
\label{sec:metr-cell-resampl}
As described in the introduction, the Cell Resampler will narrow the distribution of weights of the events by first forming clusters of \emph{cells} of events nearby to the negative events, and then assigning each event in the cells the weight of the average of weights in the cell.
Starting from the most negative event, the cell is allowed to grow until the sum of weights is positive, or alternative a maximum distance from the cell seed is reached without achieving a positive sum.

This discussion presumes a measure of the distance between events.
Much of the cancellation embedded in the fixed-order perturbative calculation is between events differing only by radiation of soft or collinear origins.
One can ensure that the distance between such events is small by letting the distance depend not on the momenta of all the particles in the events, but only on infra-red safe constructs such as jets, isolated photons and dressed leptons.

In order to apply an efficient algorithm in finding nearest neighbours in a sample with many events, the distance is required to have the full set of properties of a metric.
This is achieved by having a metric for each type of $t$ particle (e.g.~jet, photon, lepton) and set the distance between two events $e,e'$ to be~\cite{Andersen:2023cku}
\begin{equation}
  \label{eq:d_event}
  d(e,e') = \sum_t d(s_t, s_t'),
\end{equation}
where $d(s_t, s_t')$ is the distance between the two sub-sets of particles $s_t, s_t'$.
With an IR-safe definition of jet flavour, one could extend the metric by restricting each set to be of the same jet flavour.
Similarly, each lepton flavour and charge should be considered separately.

Finally, let us discuss a good definition of the distance between two sets of multiple identical IR-safe objects.
If the set of particle type $t$ contains just one element in each event then the distance $d(s_t, s_t')$ could be just the difference between the three-momenta in $s_t$ and $s_t'$.
When each set contains multiple momenta (restricting for now the discussion to the case of the same multiplicity in each set), one could think of ordering both according to energy, or transverse momenta etc., and sum the distance between each set of ordered elements.
However, this is often not the comparison which would result in the smallest such distance.
A better definition (meaning more resilient against small changes in the momenta) is found by minimising over the possible orderings
\begin{equation}
  \label{eq:d_set}
  d(s_t,s_t') = \min_{\sigma \in S_P} \sum_{i=1}^P d(p_i, q_{\sigma(i)})\,,
\end{equation}
where $p_1,\dots, p_P$ are the momenta of the objects in $s_t$ and $q_1,\dots, q_P$ the momenta of $s_t'$.
If the sets of particles from the two events contain a different number of elements then elements with zero-momentum entries can be inserted into the smallest set until the two sets contain the same number of elements.
The na\"ive factorial complexity of calculating the minimum over all the possible orderings between two sets of $N$ elements is avoided using the ``Hungarian method''~\cite{Jacobi1,Jacobi2,Kuhn1,Kuhn2,Munkres,10.1145/321694.321699,https://doi.org/10.1002/net.3230010206}

Finally, the distance $d(p,q)$ between two vectors is chosen as~\cite{Andersen:2021mvw}
\begin{equation}
  \label{eq:p_dist}
  d(p, q) = \sqrt{\lvert\vec{p} - \vec{q}\,\rvert^2 + \tau^2 (p_\perp - q_\perp)^2},
\end{equation}
where $p_\perp$ and $q_\perp$ are perpendicular to the beam axis and $\tau$ is an adjustable parameter.
All the fixed-order studies in Ref.~\cite{Andersen:2023cku} used $\tau=0$, such that the distance is just the length of the difference of the spatial part of the vectors.
We will in this paper discuss the significance of other choices for $\tau$.

\subsection{Photon Isolation}
\label{sec:photon-isolation}
In order to apply cell resampling to the diphoton background without modifying observables sensitive to the photons (e.g. the leading and sub-leading photon $p_T$ or diphoton $p_T$), the metric must depend on the photon momenta.
For small allowed cell radii this ensures that cell resampling does not redistribute weights between events with vastly different photon energies.
As discussed, the metric should depend only on IR safe objects to ensure that there is a small distance between events differing by soft and collinear emissions only.
The photon momenta will therefore be defined using an isolation criterion.

\begin{table}[]
\begin{tabular}{l|l|l|l}
Parameter                   & Runcard Name                      & Default & Value \\
\hline
$\epsilon_\gamma$           & \texttt{photonefrac}  & 0.09 & 0.1    \\
$R$                         & \texttt{photonradius} & 0.2 & 0.1     \\
$p_T^{\gamma_\mathrm{min}}$ & \texttt{photonpt}     & 20& 17\\
\hline
\end{tabular}
\caption{Photon isolation parameters and their corresponding \texttt{cres} runcard names, default values, and values used in the current study.}
\label{tab:photon_isolation}
\end{table}

For this work, we have implemented a fixed cone photon isolation algorithm, as used in the ATLAS and CMS experimental measurements.
It is straightforward to extend the algorithm to apply other isolation procedures, for example, smooth cone isolation~\cite{Frixione:1998jh}.
When constructing a list of isolated photons, the cell resampler first checks if a photon, $\gamma_i$, is isolated by requiring
\begin{align}
&p_T^{\gamma_i} \ge p_T^{\gamma_\mathrm{min}}, &
&E_T^{\gamma_i} \ge \epsilon_\gamma E_T^\mathrm{cone}(R). &
\end{align}
Here, $p_T^{\gamma_i} $ is the transverse momentum of the photon and $p_T^{\gamma_\mathrm{min}}$ is the minimum $p_T$ required to consider it isolated.
The quantity $E_T^\mathrm{cone}(R)$ is the transverse energy of all particles (excluding neutrinos and muons) entering a cone of radius $R$ centred around the photon, $E_T^{\gamma_i}$ is the transverse energy of the photon and $\epsilon_\gamma$ is the minimum fraction of the energy of the cone that the photon must carry to be considered isolated.
A particle is inside the cone if $r \le R$ with $r=\sqrt{(\Delta \eta)^2 + (\Delta \phi)^2}$, where $\Delta \eta$ and $\Delta \phi$ are the pseudorapidity and azimuthal angle separation between the photon and the particle.
The fixed cone isolation parameters can be set using the cell resampler configuration file, see Table~\ref{tab:photon_isolation} for the corresponding flags and the values used in this study.
These are similar to the values used in the generation of the event samples.

\section{Results}
\label{sec:results}
In order to investigate the impact of the various steps in the generation on the efficiency of the cell resampler we begin by generating a large set of weighted and unweighted $p p \rightarrow \gamma \gamma + \mathrm{jets}$ events at perturbative orders which excludes Higgs boson production. These events are then relevant for the study of the Standard Model background to Higgs boson production.
Specifically, we choose to follow the analysis of Ref.~\cite{ATLAS:2021mbt}, and the histograms shown later are generated using the corresponding Rivet~\cite{Bierlich:2019rhm} analysis.
We have furthermore implemented analyses of the momenta of any jets in the event (their transverse momentum, rapidity distributions, dijet invariant masses etc.).
In the following we will pick distributions to highlight specific effects and mention their generality.

Some of the cuts applied in the analysis influence the generation cuts, meaning that just as in the case of fixed order analyses, the generation cuts need to be tailored to the analysis in order to ensure a correct description.
In particular, the transverse momentum of the hardest isolated photon is required to be larger than 40~GeV, while that of the second hardest must be larger than 30~GeV.
It is therefore clear that a precise description of the emission of partons with $p_t>10$~GeV is necessary and should be performed at least at Born level accuracy (instead of shower accuracy).
As discussed, we will investigate samples of shower merged samples of increasing multiplicity of jets, each of NLO+shower accuracy.

We will study in turn the following event samples:
\begin{enumerate}[A.]
\item MadGraph+Pythia showered and multi-jet (0,1,2@NLO) merged unweighted diphoton sample.\label{item:1}
\item Sherpa showered and multi-jet (0,1,2@NLO) merged weighted diphoton sample.\label{item:3}
\end{enumerate}

To illustrate the effects of the Cell Resampling, we will use just a few examples which capture the behaviour seen in the myriad of cases studied:
\begin{enumerate}
\item $d\sigma/dp_{t,\gamma_1}$: the distribution on the transverse momentum of the hardest photon.
\label{item:2}
\item $d\sigma/d\theta_{\mathrm{cs}}$: the distribution on the polar scattering angle in the Collins-Soper rest frame~\cite{Collins:1977iv}.
\label{item:4}
\item $d\sigma/dp_{t,\gamma\gamma}$: The transverse momentum of the di-photon system.
\label{item:5}
\end{enumerate}

The reasoning for presenting these distributions is as follows:
The first distribution uses a single momentum, which itself enters the metric directly.
With a well-defined minimum value (at a perturbative scale chosen by the cut) and bin widths increasing from roughly 5~GeV, it is perhaps not surprising that the results for this observable are well behaved.
The polar angle in the Collins-Soper frame is a quantity formed from the photon momenta which enter the metric.
However, the relation is non-linear, and the binning width has no units and cannot be related to cells of a specific size.
This quantity is of interest for this reason, and for others which will be revealed later.
Finally, the distribution on the transverse momentum of the diphoton system is interesting, since the sum of the photon momenta does not itself enter the metric, and the measurement extends to $2.5$~GeV with a bin width of just a few GeV at the low end of the distribution.
This distribution is therefore a severe stress test of the metric.

We will observe that applying the cell resampler with the same metric ($\tau=0$ in Eq.~\eqref{eq:p_dist}) which resulted in orders of magnitude improvements for the fixed order calculations in Ref.~\cite{Andersen:2023cku} leads to almost no reduction in the contribution from negative weights.
As we will show, this is due to two reasons.
Firstly, the negative weight contribution in the shower merged samples are less severe than in the fixed order samples investigated so far.
There is therefore less of a potential for improvements.
Secondly, using the metric with $\tau = 0$, the shower merged samples have the events with large negative weights so isolated from events with positive weights that the cell size would have to grow too big to organise cancellations to not also distort distributions.
This is true for both types of shower merging investigated, and is different to the behaviour observed in fixed order calculations, where the large cancellations are mostly of infra-red origin and therefore local in the $\tau = 0$ metric.

However, inspired by a cursory observation of the bin-to-bin fluctuations we find that choosing $\tau>0$ in the metric allows for large cancellations with distortions less than the statistical uncertainty (and less than 1\% in each bin) in the distributions~\ref{item:2}-\ref{item:4} discussed above.
The study of an adaptive metric which accounts properly for distributions in derived momenta approaching small scales is left for a future publication.

\subsection{Showered \& Multi-jet Merged $\gamma \gamma + \mathrm{jets}$  Unweighted Sample}
\label{sec:showered--multi}
We consider a set of 10M showered, multi-jet merged and unweighted events generated using \textsc{MadGraph5\_aMC@NLO}~\cite{Alwall:2014hca} interfaced with the PYTHIA~\cite{Bierlich:2022pfr} parton shower.
Each of the processes $pp\to\gamma \gamma + i\ \mathrm{jets}$ , $i=0,1,2$ are calculated to NLO accuracy in QCD.
Divergences in the radiation of photons off quarks was avoided by imposing a generation cut using Frixione's isolation~\cite{Frixione:1998jh} with parameters $\delta_0=0.1$, $n=2$ and $\epsilon_{\gamma}=0.1$.
For the generation, jets were reconstructed with the $kt$ algorithm with a jet radius of 1.0 and a minimum jet transverse momentum of 10~GeV.
The shower merging employed the FxFx scheme~\cite{Frederix:2012ps} with a merging scale of 10~GeV.
We refer to appendix~\ref{sec:choices-metric} for further discussions, but here just note that the photons entering the cell resampler metric for this analysis were found with the parameters and algorithm used also in the generation.

Before considering the effects on distributions, we list in Table~\ref{tab:aa} the contribution from negative weighted events to the total cross section of the analysis for increasing maximum cell radii obtained using the metric in Eq.~\eqref{eq:p_dist} with $\tau = 0$.
With a maximum cell radius of 0~GeV the contribution is 0.297 (0.296) in a sample with 1M (10M) events.
The fact that a procedure with even 100~GeV maximum cell radius reduces this to just 0.195 (and cannot cancel all negative events) is an indication that the phase space is full of regions where the cross section is locally negative.
\begin{table}[h!]
\centering
\begin{tabularx}{\columnwidth}{@{}>{\raggedright\arraybackslash}X|>{\raggedright\arraybackslash}X >{\raggedright\arraybackslash}X|>{\raggedright\arraybackslash}X >{\raggedright\arraybackslash}X@{}}
\hline
Max Cell Size& $r_-^\mathrm{1M}$ & $r_-^\mathrm{10M}$ & $c(r_-^\mathrm{1M})$ & $c(r_-^\mathrm{10M})$  \\ \hline
0                           & 0.297                          & 0.296                           & 6.07         & 6.01   \\
10 GeV                 & 0.296                          & 0.294                           & 6.01         & 5.89   \\
30 GeV                 & 0.287                          & 0.280                           & 5.51        & 5.17   \\
60 GeV                 & 0.261                          & 0.245                           & 4.38         & 3.84   \\
100 GeV                & 0.218                          & 0.195                           & 3.14         & 2.69  \\
$\infty$                 & 0.000                          & 0.000                          & 1               & 1  \\ \hline
\end{tabularx}
\caption{The negative event fractions, $r_-^\mathrm{1M}$ and $r_-^\mathrm{10M}$, obtained using cell resampling on a set of 1M and 10M showered, multi-jet merged and unweighted $\gamma \gamma + \mathrm{jets}$ events.}
\label{tab:aa}
\end{table}

\begin{figure}
  \centering
    \subfigure[]{\includegraphics[width=\columnwidth]{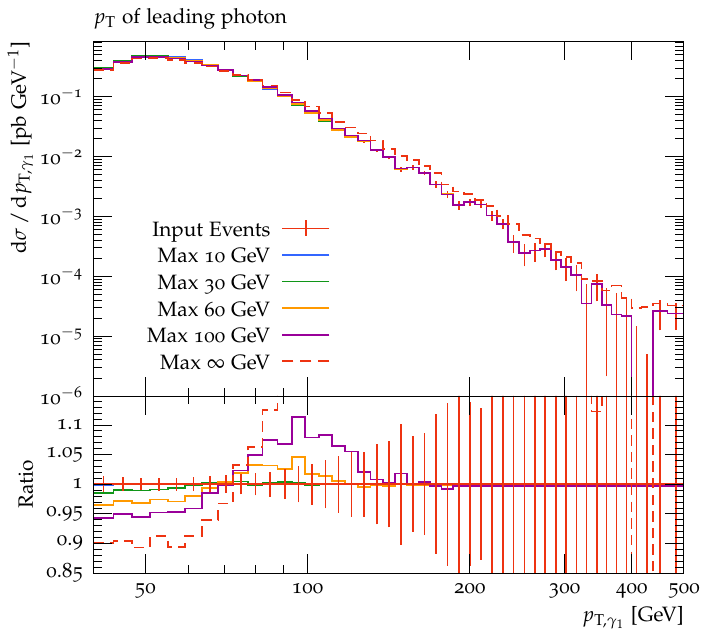}}
    \subfigure[]{\includegraphics[width=\columnwidth]{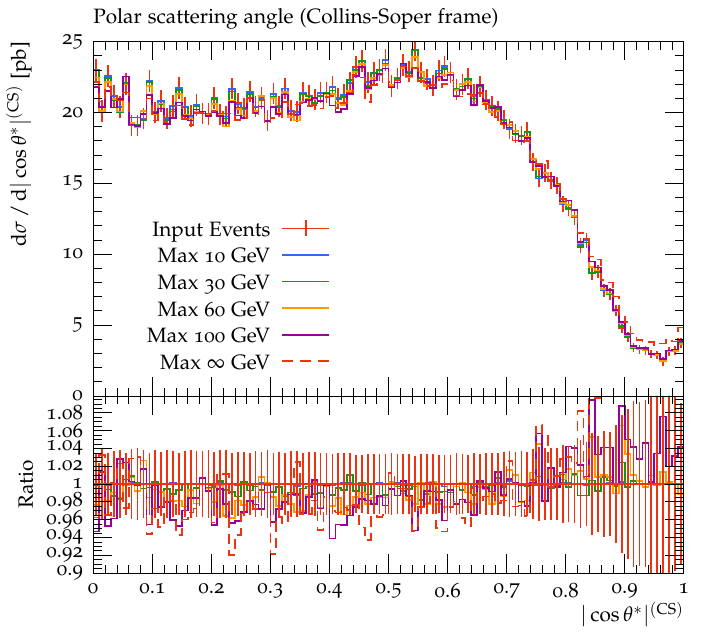}}
    \caption{Transverse momentum of the leading photon and the polar angle in
      the Collins-Soper frame for the diphoton system using the unweighted
      MC@NLO showered and FxFx multi-jet merged $\gamma\gamma$ inclusive
      sample discussed in Section~\ref{sec:showered--multi}.}
  \label{fig:showeredaa}
\end{figure}

\begin{figure}
  \centering
    \includegraphics[width=\columnwidth]{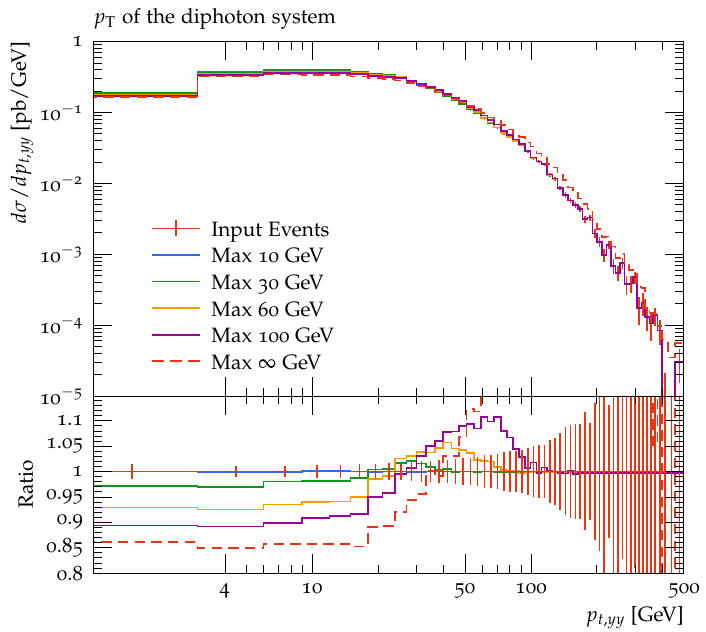}
    \caption{Transverse momentum of the diphoton
      system for the unweighted MC@NLO showered and FxFx multi-jet merged $\gamma\gamma$
      inclusive sample discussed in Section~\ref{sec:showered--multi}.}
  \label{fig:showeredaa2}
\end{figure}

The cell resampler has used every single negative weighted event as a seed for a cell which can grow to the maximum cell size, and yet even with 100~GeV maximum radii the contribution from left over negatively weighted events is reduced to only 0.195.
Such events could cause large bin-to-bin fluctuations in the histograms of the analysis.

However, such large cells in the standard metric would cause distortions of the distributions.
We note here that for the 10M event sample the median cell radius is $92.5$~GeV when run with unrestricted cell size (and obtaining complete cancellation of negative events).
Increasing the size of the sample from 1M to 10M events reduces the median cell radius for complete cancellation from $112.4$~GeV to $92.5$~GeV.
The cells really need this large a radius for the local cross section in the generated sample to be positive.
The reduction in the median cell radius from 1M to 10M indicates that increasing the density of events in phase space aids the local cancellation of negative weights.

The results for the kinematic distributions introduced in Sec.~\ref{sec:results} are presented in Figs.~\ref{fig:showeredaa}-\ref{fig:showeredaa2}.
The distributions resulting from the input events, prior to cell resampling, are shown in red, with the Monte Carlo uncertainty indicated by the error bars.
The transverse momentum distribution of the hardest photon (Fig.~\ref{fig:showeredaa}(a)) illustrates very clearly the convergence of the result from the cell resampler to that of the original sample as the maximum allowed cell size is decreased.
The results with 10~GeV maximum cell size are indistinguishable from the input sample.
Even a 30~GeV max cell size results in deviations only just outside 1$\sigma$ statistical significance and less than 2\% (but systematically undershooting for $p_{T,\gamma_1}<60$~GeV).
As discussed previously, the convergence is expected, since the photon momentum directly enters the metric.
We note that the size of the discrepancy between the cell resampled and original results is a function of both the bin width and of the gradient of the distribution.
In particular, with a fixed maximum cell size, the distortions tend to be larger for steeply falling distributions in regions with small bin widths.
This is especially visible for $p_{T,\gamma_1}$ around 100~GeV when the maximum cell size is 100~GeV or greater.
Conversely, at small $p_{T,\gamma_1}$ which has smaller bin widths, the deviations are smaller even with a large maximum cell size due to the flatness of the distribution.

Fig.~\ref{fig:showeredaa}(b) shows the distribution on the polar scattering angle in the Collins-Soper frame.
This scattering angle has a non-trivial dependence on the momentum of the two photons.
However, we observe that the analysis of the cell resampled events reproduce that of the input events within the statistical uncertainty even for large maximum cell sizes.
This observation will inspire the adjustments to the metric we present in Section~\ref{sec:taunot0}.

The results for the transverse momentum of the diphoton system are shown in Fig.~\ref{fig:showeredaa2}.
The performance of the cell resampler for this distribution is interesting, as the observable $p_{t,yy}$ also does not directly enter the metric.
Nevertheless, the convergence of the CRES output to the original sample is clearly illustrated for decreasing maximum cell sizes.
The result obtained with 10~GeV maximum cell size is again almost identical to the input (the corresponding curve is hidden under that of the input results).
This is of interest here because the scale on the plot extends to scales below 4~GeV and bin widths even smaller.
The results obtained with a 30~GeV maximum cell size are here undershooting systematically by 3\% for scales below 20~GeV.

It is clear that in general the applied metric is suboptimal for the set of shower-merged NLO-matched events, given that the negatively weighted events are often isolated using this metric.
This is in contrast to standard next-to-leading order events with cancellations dominated by infra-red connected events and counter-events.

The results for an alternative shower-merging will now be investigated.

\subsection{Showered \& Multi-jet Merged $\gamma \gamma + \mathrm{jets}$  Weighted Sample}
\label{sec:showered--multi-1}

We next investigate the situation with a sample of weighted events generated using a different multi-jet merging algorithm. Specifically, we investigate events generated with the Sherpa~\cite{Sherpa:2019gpd} MEPS@NLO~\cite{Gehrmann:2012yg} method, merging $pp\to\gamma\gamma+0,1,2$j samples, each shower sample matched to NLO using the MC@NLO~\cite{Frixione:2002ik} procedure.
In order to have comparable statistical uncertainties to those of the unweighted sample, a weighted sample of 100M events was generated.
The weighted events were passed through the Cell Resampler with parameters for the jet clustering chosen to be suitable for the new merging procedure and merging scale.
The jet clustering parameters must be chosen to be sufficiently resolving that the analysis of the output from the Cell Resampler agrees with the analysis of the input events as the maximum cell size tends to zero.
This would not be the case if the jet threshold was set too high, which could put all events into a 0-jet bin and average weights irrespectively of the jet multiplicity.
For fixed order analyses the jet threshold for use in the metric can be set at the analysis scale or smaller.
For the FxFx merged sample studied in section~\ref{sec:showered--multi} the jet $p_T$ threshold was set to the jet merging scale (measured in the lab frame).
For the generated MEPS@NLO sample with $Q_0=20$~GeV (not measured in the lab frame) it was found that a jet pt threshold of 5~GeV (or smaller) worked.

Table~\ref{tab:sherpa_aa_weighted} shows the ratio of the absolute sum of negative weights to the sum of absolute weights for the input and cell resampled events with various maximum cell sizes using the metric in Eq.~\eqref{eq:p_dist} with $\tau = 0$.
This quantity corresponds to the fraction of negatively weighted events that would result from unweighting of the event sample.
We begin by noting that with the current setup, the sample generated with Sherpa has fewer negative events to begin with, and therefore also less room for improvements.
We will see that for this setup, the negative events are again too isolated in phase space to obtain a sizeable cancellation without also leading to distortions of the distributions.

\begin{table}[b]
\centering
\begin{tabularx}{\columnwidth}{@{}>{\raggedright\arraybackslash}X|>{\raggedright\arraybackslash}X|>{\raggedright\arraybackslash}X>{\raggedright\arraybackslash}X|>{\raggedright\arraybackslash}X@{}|}\hline
  Max Cell Size
                                       &$r_-^\mathrm{100M}$  
  & $c(r_-^\mathrm{100M})$  \\ \hline
Input & 0.176 & 2.38 \\
0 GeV & 0.176 & 2.38  \\
10 GeV & 0.176 & 2.38   \\
20 GeV &  0.173 & 2.34 \\
25 GeV &  0.169 & 2.28 \\
30 GeV & 0.133  & 1.86 \\ \hline
\end{tabularx}
\caption{Negative weight fractions for showered weighted $\gamma \gamma$ sample.}
\label{tab:sherpa_aa_weighted}
\end{table}

\begin{figure}
  \centering
    \subfigure[]{\includegraphics[width=\columnwidth]{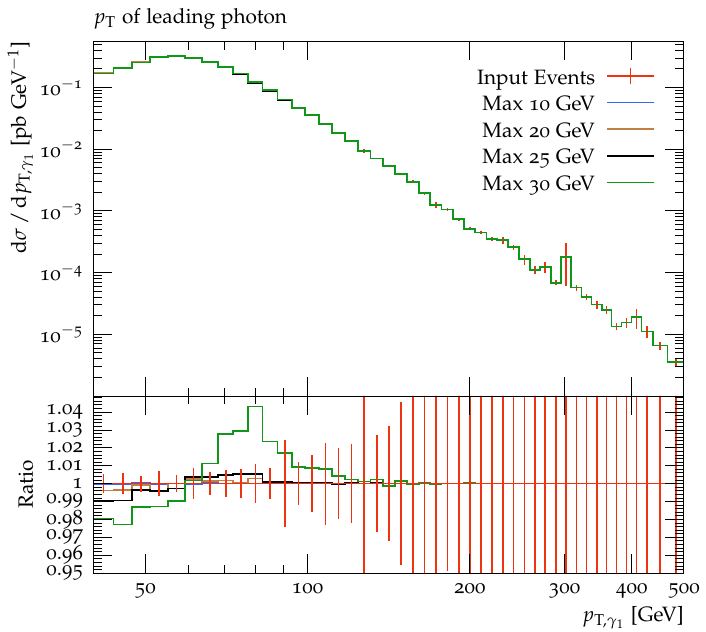}}
    \subfigure[]{\includegraphics[width=\columnwidth]{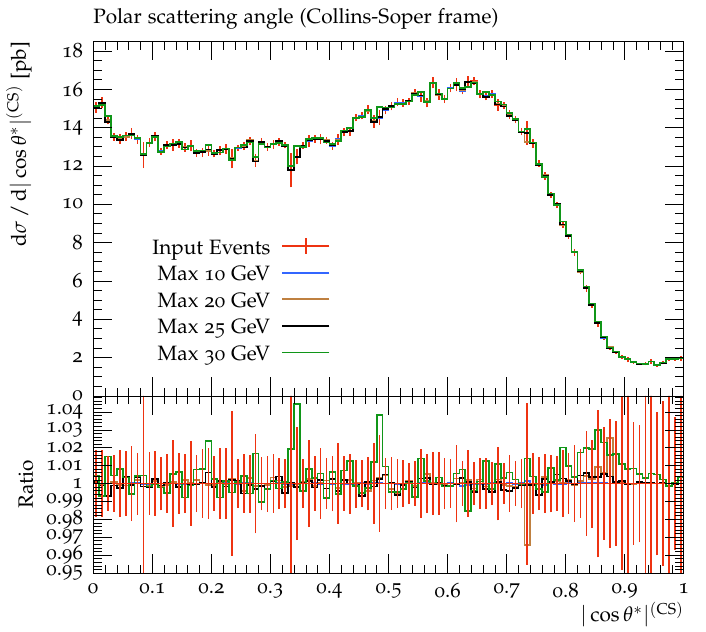}}
    \caption{Transverse momentum of the leading photon and the polar angle in
      the Collins-Soper frame for the diphoton
      system using the weighted MC@NLO showered and MEPS@NLO multi-jet merged
      sample obtained with Sherpa and discussed in Section~\ref{sec:showered--multi-1}.}
    \label{fig:dsigmadpta1}
\end{figure}

The result for the transverse momentum of the leading photon is shown in Fig.~\ref{fig:dsigmadpta1}(a).
The distributions resulting from the input events, prior to cell resampling, are shown in red.
We also display results for the distributions after resampling the events with max cell sizes of $10, 20, 25$ and $30$~GeV.
Decreasing the max cell radii causes the cell resampled distributions to converge towards the input distribution.
We observe that the resampled results with a max cell size of $10$~GeV are indistinguishable from that of the input events.
The results obtained using a maximum cell size of $20$~GeV reproduce the $p_T$ distribution of the leading photon within the Monte Carlo uncertainties.
With maximum cell sizes of $25$ and $30$~GeV we observe deviations beyond the Monte Carlo uncertainties.
The general pattern of deviations between the input and resampled distributions follows that already observed in Section~\ref{sec:showered--multi}.

Fig.~\ref{fig:dsigmadpta1}(b) shows that the distribution on the scattering angle in the Collins-Soper frame is well reproduced after cell resampling, just as was the case in Section~\ref{sec:showered--multi}.

\begin{figure}
  \centering
  \includegraphics[width=\columnwidth]{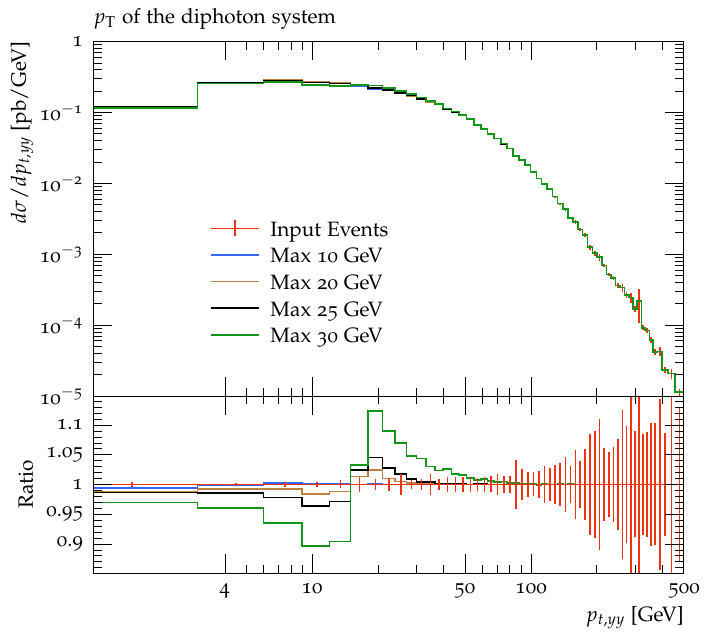}
  \caption{The distribution on transverse momentum of the diphoton system using
    the sample of Section~\ref{sec:showered--multi-1}.}
  \label{fig:aayypt}
\end{figure}
Fig~\ref{fig:aayypt} shows the transverse momentum of the diphoton system. The results from the 10~GeV max cell size reproduce the input within statistical uncertainties, but the deviations for the other distributions follow the pattern observed in section~\ref{sec:showered--multi}, and will again have the same solution.

\begin{figure}
  \centering
  \includegraphics[width=\columnwidth]{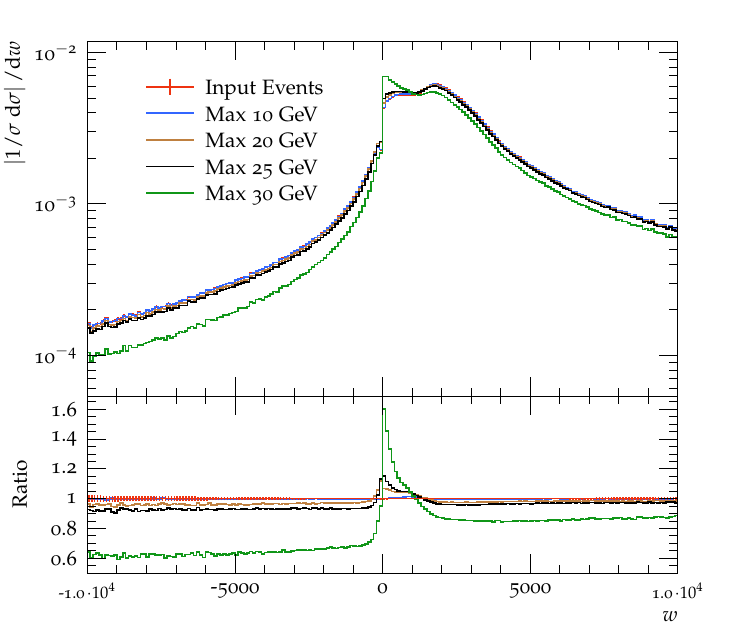}
  \caption{The weight distribution for the events in the $\gamma\gamma$
  sample discussed in Section~\ref{sec:showered--multi-1}.}
  \label{fig:aajjweights}
\end{figure}

Fig.~\ref{fig:aajjweights} shows the weight distribution for the cross section of the event sample studied here.
On this plot the left hand side has to be subtracted from the right hand side to obtain the cross section.
We observe that for the input events the distribution of event weights is smooth, with long and very significant tails to large both positive and negative weights.
With the metric used, only a few percent of the cross section in the tails is moved to the region around 0.
This of course still constitutes a reduction in the variance of the distribution, albeit only a marginal improvement for the choices leading to insignificant changes in the kinematic distributions.

\subsection{A slight modification facilitating large cancellations}
\label{sec:taunot0}

The results of the previous sections have clearly demonstrated the convergence of all results in the limit of small cell sizes.
The results have also demonstrated that within the shower-merged samples studied, the distance between the events with negative weights and events with positive weights is too large to implement any significant cancellation while keeping the modifications to distributions under control, at least when using the metric in Eq.~\eqref{eq:p_dist} with $\tau=0$.
The negative weights within the shower-merged sample are not dominated by those of infra-red nature which the metric was designed to identify as being close.
This could be a result of the procedure for merging across several jet multiplicities, which of course differ by IR safe observables (e.g.~the jet count).

However, the above observations also point towards a solution.
The cells used for cancellation clearly cannot extend far in transverse momentum without jeopardising the description.
At the same time, the effect of the long tail of negative events is cancelled in each bin of the transverse momentum distribution, and the cancellation must therefore include regions in other directions.
A simple approach to allowing such shapes of cells is to let $\tau>0$ in Eq.~\eqref{eq:p_dist}.
In this section, we present results obtained using $\tau=9$ and demonstrate that a large reduction of negative event weights can be obtained without significantly modifying observables.
The cell sizes with the $\tau>0$ metric can be numerically larger than with $\tau =0$ without extending further in the transverse direction.
In the present work, we do not attempt to optimise (or study) the choice of $\tau$, nor do we investigate other metrics, instead we defer such studies to a future work on a new class of metrics. 

In Table~\ref{tab:sherpa_aa_weighted_ptweight9} we list the negative event fractions for the cell resampled events of Section~\ref{sec:showered--multi-1} obtained using the $\tau=9$ metric.
As an illustration, with maximum cell sizes of 100~GeV the negative event weight contribution is reduced from 0.176 in the input sample to 0.107 after cell resampling.
This corresponds to a reduction from 238\% to 162\% in the number of events needed to represent the sample compared to a sample with only positive weights.
Put another way, the fraction of vexatious events could be reduced from 1.38 to 0.62 of the idealistic sample of pure positive weights.
The cell resampler obtains a median cell radius of 36.2~GeV for the maximum allowed cell size of 100~GeV.
\begin{table}[htb]
\centering
\begin{tabularx}{\columnwidth}{@{}>{\raggedright\arraybackslash}X|>{\raggedright\arraybackslash}X|>{\raggedright\arraybackslash}X>{\raggedright\arraybackslash}X|>{\raggedright\arraybackslash}X@{}|}\hline
  Max Cell Size
                                       &$r_-^\mathrm{100M}$  
  & $c(r_-^\mathrm{100M})$  \\ \hline
Input & 0.176 & 2.38 \\
30 GeV & 0.171 & 2.31   \\
60 GeV &  0.138 & 1.91 \\
100 GeV & 0.107  & 1.62 \\ \hline
\end{tabularx}
\caption{Negative weight fractions for showered weighted $\gamma \gamma$
  sample using a metric with $\tau$=9.}
\label{tab:sherpa_aa_weighted_ptweight9}
\end{table}

\begin{figure}
  \centering
    \subfigure[]{\includegraphics[width=\columnwidth]{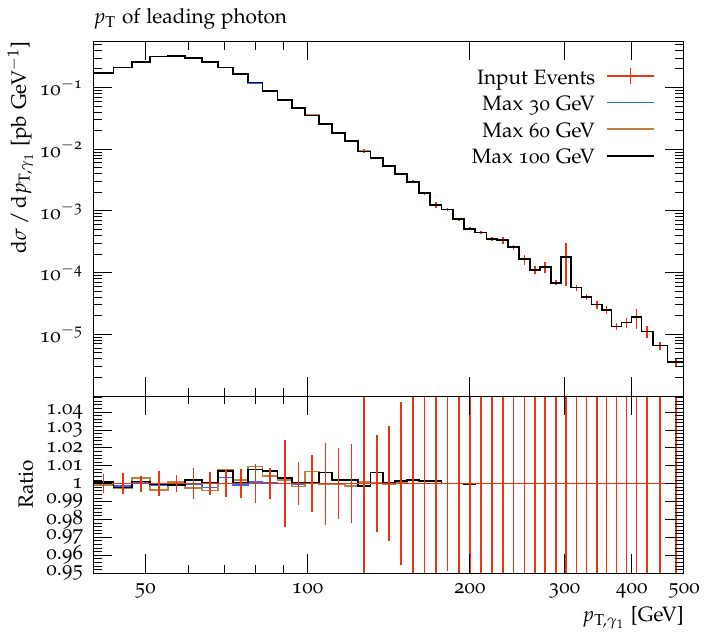}}
    \subfigure[]{\includegraphics[width=\columnwidth]{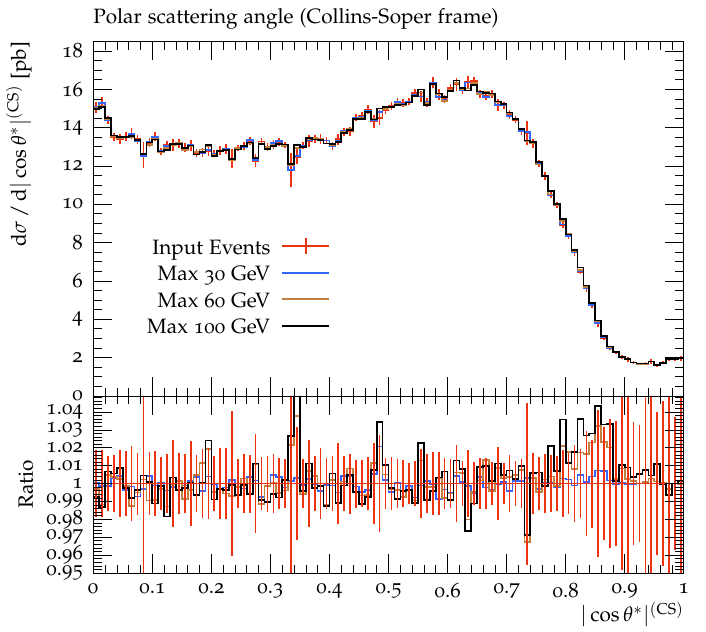}}
    \caption{
    Transverse momentum of the leading photon and the polar angle in the Collins-Soper frame for the diphoton system using the weighted MC@NLO showered and MEPS@NLO multi-jet merged sample obtained with Sherpa in Section~\ref{sec:showered--multi-1} resampled with $\tau=9$.
    These results can be directly compared to Fig.~\ref{fig:dsigmadpta1}, note that due to the different distance metrics used the meaning of maximum cell size is not equivalent.
    }
    \label{fig:dsigmadpta1ptweight9}
\end{figure}

We present in Fig.~\ref{fig:dsigmadpta1ptweight9}(a) the transverse momentum distribution of the hardest photon for the same event sample studied in section~\ref{sec:showered--multi-1}.
The results can therefore be compared to those of Fig.~\ref{fig:dsigmadpta1}(a).
All the listed choices of maximum cell sizes for this value of $\tau$ leads to cell resampled results agreeing with the results of the input distribution within the statistical uncertainty (and within 1\% in all bins).
Fig.~\ref{fig:dsigmadpta1ptweight9}(b) shows the results for the scattering angle in the Collins-Soper frame for the same choices of maximum cell sizes.
Within the uncertainty, the results are all equivalent to the input distribution, much unchanged from what was observed for $\tau=0$ in Fig.~\ref{fig:showeredaa}(b) and Fig.~\ref{fig:dsigmadpta1}(b).
We have investigated many other distributions, including the transverse component of the jet momentum (which itself enters the metric and therefore must converge for small cell sizes), the diphoton invariant mass and the dijet invariant masses (of first and second, first and third etc).
All such distributions are described well in the regions investigated using both the $\tau=0$ and $\tau=9$ metrics.

\begin{figure}
  \centering
    \includegraphics[width=\columnwidth]{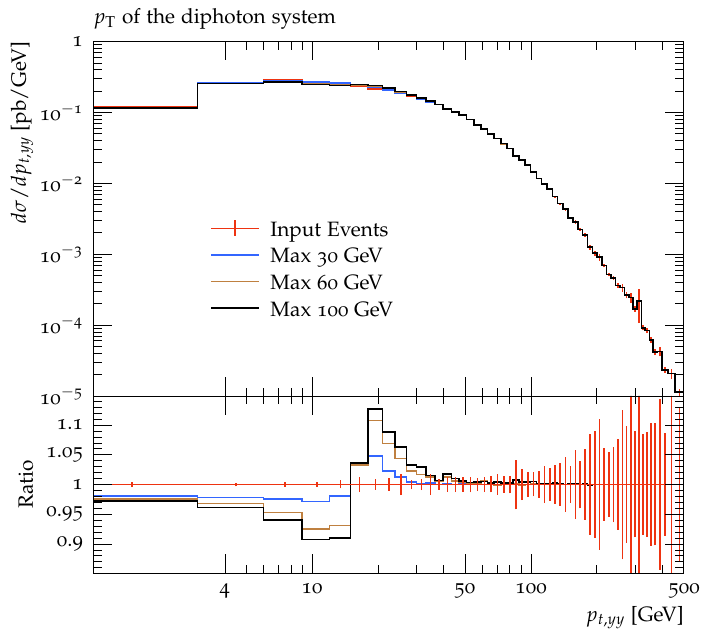}
    \caption{
    The distribution on transverse momentum of the diphoton system using the sample of Section~\ref{sec:showered--multi-1} resampled with $\tau=9$.
    These results can be directly compared to Fig.~\ref{fig:aayypt}.
    }
   \label{fig:aayyptptweight9}
\end{figure}
Fig.~\ref{fig:aayyptptweight9} shows that the issue with the description of the diphoton transverse momentum is still present at scales smaller than or similar to the maximum cell size.
This, however, can be straightforwardly solved by adding the sum of the photon momenta to the list of momenta in the metric.
One could indeed add all the combinations of perturbative momenta identified to the set of momenta considered by the metric.
The Hungarian method then becomes particularly important in calculating the minimisation over all permutations.
Work on such additions will be presented in a future work investigating a new class of metrics.

\begin{figure}
  \centering
  \includegraphics[width=\columnwidth]{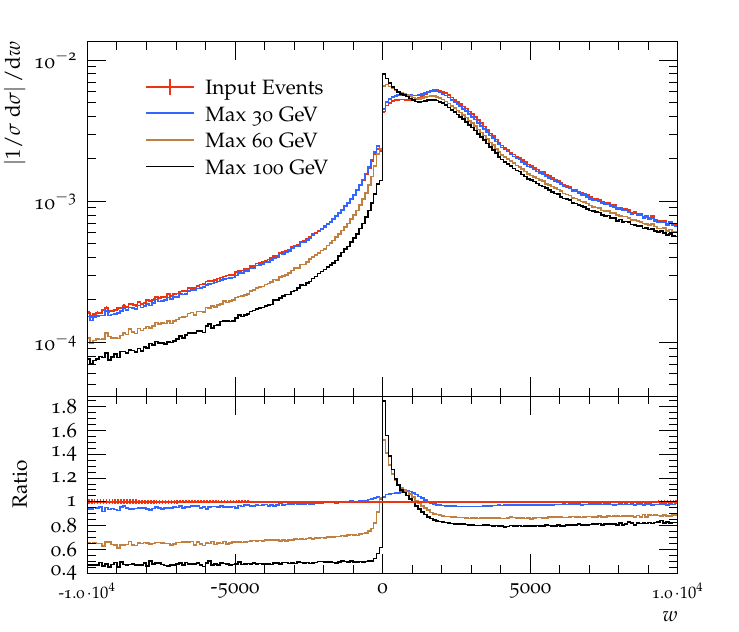}
  \caption{
  The weight distribution for the events in the $\gamma\gamma$ sample discussed in Section~\ref{sec:showered--multi-1} resampled with $\tau=9$.
  These results can be directly compared to Fig.~\ref{fig:aajjweights}.
  }
  \label{fig:aajjweightsptweight9}
\end{figure}

Finally, Fig.~\ref{fig:aajjweightsptweight9} illustrates the change in the distribution of the cross section on the weights.
The contribution to the cross section of events with a large negative weight is reduced to 40\% of that in the input sample by cell resampling with a maximum cell size of 100~GeV.

\section{Conclusions}
\label{sec:conclusions}
We have investigated the use of cell resampling and the impact of the choice of metric on events generated with two shower-merging algorithms of NLO-matched showered events.
We found that, in multi-jet merged samples, events with large negative weights were too far from positive events when resampled with the metric used to obtain an efficient cancellation in the fixed-order samples studied previously~\cite{Andersen:2023cku}.
In fixed-order samples, the origin of the cancellation is of infra-red nature and cancellation can therefore be organised locally even with the simplest metric on the space of events.
Instead, in multi-jet merged samples, we observe that negative events can be separated from positive events by a large distance in the space of 3-momenta while remaining relatively close in the plane transverse to the beam axis.
A simple modification of the metric, anticipated in Ref.~\cite{Andersen:2021mvw}, to assign a larger distance when events differ in the transverse plane allows for large cancellations of negative event weights to be obtained with the cell resampler even for the shower-merged NLO-matched events.
With the suitably defined metric, we found that the cell resampler is able to significantly reduce the fraction of negatively weighted events with almost all kinematic distributions reproduced within the statistical uncertainty and with distortions at or below the $1\%$ level.
We identified that the description of observables sensitive to derived momentum (e.g.~the diphoton system) are most affected by resampling and can still receive deviations of up to 10\%.
We anticipate that this can be avoided by adding also the derived momenta to the metric, but, defer such studies to a future investigation of a new class of metrics which incorporate the advances made here and also address this shortcoming.

\begin{acknowledgments}
  We would like to thank the Les Houches workshop ``Physics at TeV Colliders'' for a stimulating atmosphere inspiring this study.
  This work was performed in part at Aspen Center for Physics, which is supported by National Science Foundation grant PHY-2210452.
The work of Jeppe R.~Andersen and Stephen Jones is supported by the STFC under grant ST/X003167/1.
Stephen Jones is additionally supported by a Royal Society University Research Fellowship (Grant URF/R1/201268).
This work was supported in part by the Spanish Ministry of Science and Innovation (PID2020-112965GB-I00,PID2023- 146142NB-I00), the Departament de Recerca i
Universities from Generalitat de Catalunya to the Grup de Recerca 00649 (Codi: 2021 SGR 00649) and the Spanish Agencia Estatal de Investigaci\'on (AEI) under grant agreement RYC2021-031273-I funded by MCIN/AEI/10.13039/501100011033 and the "European Union NextGenerationEU/PRTR".
This project has received funding from the European Union's Horizon 2020 research and innovation programme under grant agreement No 824093.
IFAE is partially funded by the CERCA program of the Generalitat de Catalunya.
\end{acknowledgments}

\appendix
\section{Choices of perturbative objects the metric}
\label{sec:choices-metric}
We discussed in Section~\ref{sec:photon-isolation} the photon isolation procedure, which allows for an IR safe inclusion of photons into the metric.
Previous studies with the Cell Resampler have focused on fixed-order partonic calculations, where the generation and analysis criteria for jets etc.~are correlated.
There is therefore no ambiguity as to which jet parameters to choose for the IR safe objects which should enter the metric.
The situation could be slightly different in the case of showered multi-jet merged calculations, if a result depends significantly on the shower merging scale used, and the jet merging parameters are much different to any reasonable choice for the analysis.
In the current study we have used jet definitions with the most discriminating parameters of the generation and analysis for the objects entering the metric used in the Cell Resampler.
This means anti-$k_t$-jets with $R=0.4$ and a $p_t$ threshold of 10~GeV.
The jet generation cuts used for the FxFx merging discussed in Section~\ref{sec:showered--multi} are based on the $k_t$-algorithm with $R=1.0$.

If deemed necessary, one could avoid concerns about whether to choose generation or analysis parameters for the perturbative objects used in the metric, by simply having both generation and analysis objects as separate sets in the metric.
Such a choice could also be used to ensure the correct description of not just observables based on e.g.~the jet momenta, but also of e.g.~jet profiles.
This could be achieved by including sets of jet momenta of varying jet radii.

\bibliography{main.bib}

\end{document}